\documentstyle[12pt]{article}
\begin{titlepage}
\title{The $\eta$-Invariant as a Lagrangian of a Topological Quantum
Field
Theory}

\author{Ulrich Bunke\thanks{Humboldt-Universit\"at zu Berlin,
Institut f\"ur
Reine Mathematik (SFB288), Ziegelstr. 13a,  Berlin 10099.
E-mail:ubunke@mathematik.hu-berlin.de
}}
\end{titlepage}

\begin{document}

\newcommand{\R}{{\bf R}}
\newcommand{\Z}{{\bf Z}}
\newcommand{\C}{{\bf C}}

\maketitle
\newtheorem{prop}{Proposition}[section]
\newtheorem{lem}[prop]{Lemma}
\newtheorem{ddd}[prop]{Definition}
\newtheorem{theorem}[prop]{Theorem}
\newtheorem{kor}[prop]{Corollary}
\newtheorem{ass}[prop]{Assumption}
\newtheorem{con}[prop]{Conjecture}
\newtheorem{axiom}[prop]{Axiom}

\tableofcontents
\section{The category}

A topological quantum field theory (TQFT) starts from the categories
$C$
of oriented compact $n$-dimensional
manifolds $(M,N)$ with boundary $N=\partial M$ and $\partial C$ of
closed
$(n-1)$-dimensional manifolds.
The morphisms are orientation and boundary preserving
diffeomorphisms.
Let $W$ denote the category of finite-dimensional complex Hilbert
spaces.
The morphisms in $W$ are the isometries.
Defining a TQFT is specifying a functor $H:\partial C\rightarrow W$
as well as
a section
$E$ of the composition $H\circ \partial$, where
$\partial:C\rightarrow \partial
C$ is the functor of
taking oriented boundaries.
For the convenience of the reader we recall the notion of a section
of a
functor.
\begin{ddd} Let $F:A\rightarrow B$ be a functor between the
categories $A,B$.
A section $E\in \Gamma(F)$ of $F$ associates to any object $X\in A$
an element
$E(X)\in F(X)$ such that
for any morphism $a:X\rightarrow Y$ in $A$ we have $F(a)E(X)=E(Y)$.
\end{ddd}
The notion of a section of a functor is similar to the definition of
a parallel section of a flat vector bundle.
$H$ associates functorially to any oriented closed
$(n-1)$-dimensional manifold
$N\in \partial C$ a finite-dimensional complex Hilbert spaces $H(N)$.
One requires that $H$ satisfies the following axioms.
\begin{axiom}[orientation]\label{orrrrr} $H(-N)=H(N)^v$\end{axiom}
Here $V^v$ denotes the dual space of $V$ and $-N$ is $N$ with the
opposite
orientation. One requires that the equality in the axiom is
functorial,
i.e. for an orientation preserving diffeomorphism $f:N_0\rightarrow
N_1$
we have
$$\begin{array}{ccc} H(-N_0)&=&H(N_0)^v\\
                      H(F)\downarrow&& H(F)^v\uparrow\\
                     H(-N_1)&=&H(N_1)^v,
\end{array}$$
where the horizontal equalities are given by the axiom.
\begin{axiom}[additivity] $H(N_0\oplus N_1)=H(N_0)\otimes H(N_1)$.
\end{axiom}
Here $N_0\oplus N_1$ denotes the disjoint union of manifolds.
Again this identifiction is required to be
compatible with Axion \ref{orrrrr} and to be functorial, i.e. for
orientation
preserving
diffeomorphisms $f:N_0\rightarrow\bar{N_0}$ and $g:N_1\rightarrow
\bar{N_1}$
$$\begin{array}{ccc} H(N_0\oplus N_1)&=&H(N_0)\otimes H(N_1)\\
                     H(f,g)\downarrow&& H(f)\otimes H(g)\downarrow\\
                     H(\bar{N_0}\oplus
\bar{N_1})&=&H(\bar{N_0})\otimes
H(\bar{N_1}).
\end{array}$$
By convention $H(\emptyset)$ is canonically identified with $\C$.

The section $E$ associates to every compact oriented $n$-dimensional
manifold
$(M,N)\in C$ with boundary
$N$ a vector $E(M,N)\in H(N)$ satisfying the orientation, additivity
and the
locality
axioms.
\begin{axiom}[orientation] $$Tr E(-M,-N)\otimes
E(M,N)=1,$$\end{axiom}
where $Tr:H(-N)\otimes H(N)\rightarrow \C$ is the pairing given by
Axiom
\ref{orrrrr}.
\begin{axiom}[additivity] $$E((M_0,N_0)\oplus
(M_1,N_1))=E(M_0,N_0)\otimes
E(M_1,N_1)$$ \end{axiom}
Here the vectors are compared using the identification $H(N_0\oplus
H_1)=H(N_0)\otimes H(N_1)$.
In order to state the locality axiom note that if two compact
oriented
manifolds $M_0,M_1$
with boundary have a boundary component $N$, $-N$, respectively (here
we
fix the identifications $N\rightarrow M_i$, $i=0,1$), we can glue
$M_0$ and
$M_1$ along $N$
obtaining a new compact oriented manifold manifold
$M:=M_0\sharp_N M_1$. The glueing does not affect the other boundary
components.
Thus, the boundary of $M$ consists of the union of the boundary
components of
$M_0$ and $M_1$ different from $-N,N$.
Using the orientation axiom and the additivity axiom for $H$
we get a natural map
$$Tr: H(\partial M_0)\otimes H(\partial M_1)\rightarrow H(\partial
M)$$, which
contracts the
pair $H(N)\otimes H(-N)$ inside $H(\partial M_0)\otimes H(\partial
M_1)$. We
will not always use the pair notation $(M,N)$, in particular
if more than one boundary component is involved.
\begin{axiom}[locality] $$E(M_0\sharp_N M_1)=Tr \: E(M_0)\otimes
E(M_1)$$
\end{axiom}
The use of that functorial language is very convenient since we have
to take
into account very carefully the automorphisms of the objects
involved.
Indeed, there will be a great difference between an isomorphism and a
canonical
isomorphism. All the theory is based on this fact.

The axioms stated above are the Atiyah-Segal-Axioms of a TQFT
\cite{atiyah89},
see also the papers of Freed \cite{freed931} and Freed/Quinn
\cite{freedquinn93}.
Of special interest are the values of $E$ on closed manifolds,
which are complex numbers. By their very definition they are
diffeomorphism
invariants.

We are going to use the $\eta$-invariant of twisted signature
operators in
order to define
a local Lagrangian of a TQFT. Since the $\eta$-invariant depends
heavily on the
choice of Riemannian metrics we will
use a difference construction.
By the local variation formula the metric dependence will drop out.
This involves flat bundles, which therefore have to be included into
the
category.
Thus, we consider the category $D$ consisting of compact oriented
$n$-dimensional manifolds $(M,N,F)$
with boundary $N$ equipped with a flat Hermitian vector bundle $F$.
Analogoulsy, $\partial D$ consists of pairs $(N,F_N)$, where $N$ is
an
$(n-1)$-dimensional closed oriented manifold
and $F_N$ is a flat Hermitian vector bundle over $N$. The morphisms
in both
categories
are
pairs $(f,\Phi)$ of orientation preserving
diffeomorphisms $f$ and isomorphisms $\Phi$ of flat Hermitian bundles
over $f$,
i.e.
$$\begin{array}{ccc} F_0&\stackrel{\Phi}{\rightarrow}&F_1\\
                     \downarrow&&\downarrow\\
                     M_0&\stackrel{f}{\rightarrow} &M_1
\end{array}.$$
A local Lagrangian of a TQFT looks similar to a TQFT itself. It is
given by a
functor $H:\partial D\rightarrow V$ and a section $E\in
\Gamma(H\circ\partial)$
satisfying
the orientation, additivity and locality axioms stated above
(modified in the
obvious way for $D$, $\partial D$).
Note that the local Lagrangian of a TQFT, we will define, satisfies
the
orientation axiom up to a factor
$\pm 1$. This can be avoided by squaring everything.

\section{The $\eta$-invariant of twisted signature operators}

In this section we describe the $\eta$-invariant of twisted signature
operators
on oriented Riemannian manifolds of dimension $n=4k-1$ with boundary.
It depends on the choices of a boundary condition and a Riemannian
metric.
One result of \cite{bunke936} was to clarify the dependence of the
$\eta$-invariant
on these choices, in particular on the boundary condition. These
results will
later be used to make things independent of
the boundary condition and the choice of metrics.

Let $(M,N)$ be an $n=4k-1$-dimensional Riemannian manifold with
boundary.
We always assume that the Riemannian metric $g$ of $M$ has a product
structure
near
$N$, i.e. $g$ is of the form $g=(dr)^2+g^N$, where $r$ is the outer
normal
coordinate
to $N$ and $g^N$ is a Riemanian metric on $N$ independent of $r$.

Let $F\rightarrow M$ be a flat Hermitian vector bundle over $M$.
Then we can consider the twisted signature operator $D_F$ acting
on sections of $\Lambda^{ev}T^v M\otimes F$. Let $\omega\otimes f$
be a local section of that bundle with $\nabla^Ff=0$ and $\omega$ a
$p$-form,
then $$D_F(\omega\otimes f)=(-1)^{p+k}((\ast d-d\ast)\omega)\otimes
f.$$
If $F$ is trivial one-dimensional we denote the corresponding
operator,
called the odd signature operator in \cite{bunke936},
by $D$. In a neighbourhood of the boundary $N$ of $M$ the operator
$D_F$ has the structure
$$D_F=I(\frac{\partial}{\partial r} + D_{F_N}),$$ where we consider
the
sections of
$\Lambda^{ev}T^vM\otimes F$ as $r$-dependent sections of the
restriction of
that bundle to $N$
and $D_{F_N}$ is an elliptic differential operator acting on sections
of that
restriction.
$I$ is a bundle endomorphism of $(\Lambda^{ev}T^v M\otimes F)_{|N}$
satisfying $I^\ast=-I$, $I^2=-1$ and $ID_{F_N}+D_{F_N}I=0$.

The operator $D_F$ is symmetric on the sections with compact support
in the
interior of $M$. In order to define a self-adjoint extension of $D_F$
we have
to choose
a suitable boundary condition. We will use a boundary condition of
Atiyah-Patodi-Singer type
that will depend on the choice of a Lagrangian subspace $L$ in
$V:=ker(D_{F_N})$.
Note that $I$ acts on $V$.
A subspace $L\subset V$ is a Lagrangian subspace if $L\oplus IL=V$
and the sum
is orthogonal with respect to the scalar product induced by the
$L^2$-metric.
Let $pr_L$ be the projection
onto $L$ and $E_{D_{F_N}}(.)$ be the spectral family of the unique
self-adjoint
extension of $D_{F_N}$.
We define the initial domain of $D_{F,L}$ as
$$dom\:D_{F,L}=\{\psi\in C^\infty(M,\Lambda^{ev}T^\ast M\otimes
F)\:|\:\left(E_{D_{F_N}}(-\infty,0)+(1-pr_L)\right)\psi_{|N}=0\}.$$
Then $D_{F,L}$ is essentially self-adjoint and we denote the unique
self-adjoint
extension by the same symbol.
We define the $\eta$-invariant of $D_{F,L}$ to be the real number
$$\eta(M,N,F,L):=\frac{1}{\pi}\int_0^\infty Tr
D_{F,L}e^{-tD^2_{F,L}}t^{-1/2}
dt.$$
The integral converges by the results of Douglas/Wojciechowski
\cite{douglaswojciechowski91}.
Equivalently, one could
define the $\eta$-invariant as the value of the analytic continuation
of the
$\eta$-function $$\eta(s):=Tr \frac{sign(D)}{|D|^s}, \quad Re(s)>n$$
at $s=0$.
We will also use the reduced $\eta$-invariant
$$\bar{\eta}(M,N,F,L)=\eta(M,N,F,L)-dim\:ker\: D_{F,L}.$$

The $\eta$-invariant is a global spectral invariant with a local
variation
formula, i.e.
if $\delta D_F$ is a local variation of the Dirac operator given by a
variation
of the flat bundle or the Riemannian metric supported in the interior
of
$M$ the variation of the class $[\bar{\eta}(M,N,F,L)]$ modulo $2\Z$
is given by an integral
$$\delta([\bar{\eta}(M,N,F,L)])=\int_M \Omega(D_F,\delta D_F),$$
where $\Omega(D_F,\delta D_F)$ is an $n$-form defined by a
differential
polynomial in the data defining $D_F,\delta D_f$.

The reduced $\eta$-invariant depends on the metric and the choice of
the
Lagrangian $L$. Thus, it is not canonically defined for a Riemannian
manifold with boundary and a flat bundle. In the next sections we
will alter
the definition of the reduced $\eta$-invariant step by step making it
more and more canonical and independent of the additional choices.

\section{The canonical $\eta$-invariant of a manifold with boundary}

In this section we explain how $\bar{\eta}(M,N,F,L)$ can be viewed as
an element in the determinant line of the operator $D_{F_N}$.
The author has learned this idea from X. Dai and was
also inspired by the paper of Freed \cite{freed931}.
In fact, he independently found a similar construction using another
complex line.
After all, it turned out that this line, which is described in an
appendix to
the present section, is
canonically isomorphic to the determinant line.

The finite-dimensional Hilbert space $V:=ker\:D_{F_N}$
carries an
action of the involution $-\imath I$ and splits into the
corresponding
$\pm 1$ eigenspaces $V=V^+\oplus V^-$.
The determinant line of the operator $D_{F_N}$ is defined as the
one-dimensional
complex Hilbert space
$$det(D_{F_N}):=det(V^+)^v\otimes det(V^-),$$ where
$det(V^\pm)$ is the maximal alternating power of $V^\pm$.
By convention, $det(\{0\})$ is taken to be $\C$.

A Lagrangian subspace $L\subset V$ defines an element
$det(\sigma_L^+)\in det
(D_{F_N})$
that we are going to describe now.
Let $\sigma_L:=pr_L-(1-pr_L)$ be the reflection in $L$. The
involution
$\sigma_L$ anticommutes with $I$, i.e. $\sigma_L I+I \sigma_L=0$, and
splits
into $\sigma_L^\pm:V^\pm\rightarrow V^\mp$.
Then $$det(-\imath \sigma_L^+)\in
Hom(det(V^+),det(V^-))=det(D_{F,N})$$
is the element induced by $-\imath \sigma_L^+$.
In \cite{bunke936} we proved
the following result (strengthening a previous result of
Lesch/Wojciechowski
\cite{leschwojciechowski93})
about the dependence of the $\eta$-invariant on the
choice of the Lagrangian subspace $L$.
\begin{prop}
Let $L_0,L_1\subset V$ be Lagrangian subspaces. Then
$$\bar{\eta}(M,N,F,L_0)-\bar{\eta}(M,N,F,L_1)=m(IL_1,L_0)-dim(IL_1\cap
 L_0)
\:\:\:\:(mod\:\: 2\Z),$$
where the function $m$ is defined on pairs of Lagrangian subspaces by
$$m(IL_1,L_0)=-\frac{1}{\pi}\sum_{e^{\imath \lambda}\in
 spec(\frac{\imath+I}{2\imath}\sigma_{L_1}\sigma_{L_0}),\lambda\in(-\pi,\pi)}
\lambda.$$\end{prop}
The following proposition associates to the triple $(M,N,F)\in D$
equiped
with Riemannian metrics an element $e(M,N,F)\in det(D_{F_N})$.
We abbreviate
$det(D_{F_N})=:h(F,N)$.
\begin{prop}  \label{pp1}
The element $e(M,N,F):=e^{\pi \imath
\bar{\eta}(M,N,F,L)}det(-\imath\sigma_L^+)$,
defined for any Lagrangian subspace $L\subset V$, is independent of
$L$.
\end{prop}
{\it Proof:}
We will see that the $L$-dependence of $e^{\pi \imath
\bar{\eta}(M,N,F,L)}$
cancels with that
of $det(-\imath \sigma_L^+)$. In fact, using
$(\sigma_L^+)^{-1}=\sigma_L^-$ we
get
\begin{eqnarray*}
\frac{det(-\imath \sigma_{L_0}^+)}{det(-\imath
\sigma_{L_1}^+)}&=&det(\sigma_{L_1}^-)det(\sigma_{L_0}^+)\\
&=&det(\sigma_{L_1}^-\sigma_{L_0}^+)\\
&=&e^{-\imath\pi m(IL_1,L_0)+\imath \pi dim(IL_1\cap L_0)}.
\end{eqnarray*}
$\Box$\newline

Let $(f,\Phi):(M_0,N_0,F_0)\rightarrow (M_1,N_1,F_1)$ be a morphism
in the
category $D$
such that $f$ is an isometry.
Then $(\partial f,\partial \Phi):(N_0,F_0)\rightarrow (N_1,F_1)$
induces
an identification $\partial\Phi_\ast:V_0\rightarrow V_1$ compatible
with the involutions $-\imath I_j$, $j=0,1$. Thus, we obtain an
isometry
$$h(\partial \Phi,\partial
f):=det((\partial\Phi^+_\ast)^{-1})^v\otimes
det(\partial\Phi^-_\ast):h(N_0,F_0)\rightarrow h(N_1,F_1)$$
satisfying
$$h(\partial \Phi,\partial f)e(M_0,N_0,F_0)=e(M_1,N_1,F_1).$$
Moreover, $Tr\:e(-M,-N,F)\otimes e(M,N,F)=(-1)^{dim(V)/2}$.
\newline
\newline
{\bf Appendix}
\newline
\newline
We give another description of the determinant line $h(N,F)$.
Consider the finite-dimensional Hilbert space $V:=Ker(D_{F_N})$
together
with the action of the complex structure $I$. Let
$\Lambda(V,I)$ be the space (we need only the set without any
manifold
structure) of all Lagrangian subspaces of $V$.
We consider $\Lambda(V,I)$ as the set of objects of a category
$\Lambda$.
In this category there is for any pair of objects $L_0,L_1$
exactly one morphism $L_0\rightarrow L_1$.
We define a functor $K:\Lambda\rightarrow W$.
The functor $K$ associates to every object $L$ the complex line, i.e.
$K(L):=\C$.
For any morphism $L_0\rightarrow L_1$ we let $K(L_0\rightarrow L_1)$
be the
multiplication with $e^{-\pi\imath (m(IL_1,L_0)-dim(IL_1\cap L_0)}$.
Using a similar reasoning as in the proof of Proposition \ref{pp1}
one can check that this functor admits a one-dimensional space of
sections
$\Gamma(K)$. We claim $\Gamma(K)=h(N,F)$ in a canonical way.
Let $L\subset V$ be a Lagrangian subspace. Then $det(\sigma_L^+)\in
h(N,F)$
defines an identification $i_L:h(N,F)\rightarrow \C$ by
$i_L(det(\sigma_L^+))=1$.
The same Lagrangian subspace also defines an identification
$j_L:\Gamma(K)\rightarrow \C$ by evaluating the section at $L$.
For any two Lagrangian  subspaces $L_0,L_1$ by the proof of \ref{pp1}
we have $i_{L_0}\circ j_{L_0}^{-1}=i_{L_1}\circ j_{L_1}^{-1}$.
Thus, taking any Lagrangian $L\subset V$ we get a
canonical isomorphism $$j_L^{-1}\circ
i_L:h(N,F)\rightarrow\Gamma(K).$$
The motivation for this construction is that $L\subset V\mapsto
e^{\pi\imath\bar{\eta}(M,N,F,L)}$ can be viewed
as a section of the functor $K$.

\section{A difference construction}

We are going to use a difference construction in order to
avoid the metric
dependence of $e(M,N,F)$. We have introduced the flat vector bundle
$F$ into
the
construction just to do this particular step.
Let $Det$ be the functor which associates to
a closed oriented Riemannian manifold $N$ of dimension $n-1$ equipped
with a flat Hermitian vector bundle $F$ the line
$$Det(N,F):=h(N,F)\otimes h(N)^{-dim(F)}.$$
If $(f,\Phi):(N_0,F_0)\rightarrow (N_1,F_1)$ is a morphism
in $\partial D$ such that $f$ is an isometry, then we have a natural
map
$$Det(\Phi,f):=h(\Phi,f)\otimes (h(f)^{-1})^{\otimes
-dim(F)}:Det(N_0,F_0)\rightarrow Det(N_1,F_1).$$
Let now $(M,N,F)\in D$ be equiped with a Riemannian metric $g^N$ on
$N$.
We extend $g^N$ to the interior of $M$ and
set
$$\epsilon(M,N,F)(g^N):=e(M,N,F)\otimes e(M,N)^{\otimes -dim(F)}.$$
\begin{prop}
The invariant $\epsilon(M,N,F)(g^N)$ is independent of the Riemannian
metric in the interior of $M$ extending the metric $g^N$ on $N$.
\end{prop}
{\it Proof:}
Note that the the space of all metrics on $M$ extending a given
metric $g^N$ on
the boundary $N$ is linearly connected. Thus, it is enough to verify
that
$\epsilon(M,N,F)$ does not change under variations of the metric in
the
interior of $M$.
Since $F$ is flat, $D_F$ is locally
isomorphic to a direct sum of $dim(F)$ copies of $D$.
By the local variation formula we obtain
 $$\delta\epsilon(M,N,F)(g^N)=\imath\pi\epsilon(M,N,F)(g^N)
\int_M\left(\Omega(D_F,\delta D_F)-dim(F) \Omega(D,\delta
D)\right)=0.$$
$\Box$  \newline

Let $(f,\Phi):(M_0,N_0,F_0)\rightarrow (M_1,N_1,F_1)$ be a morphism
in the
category $D$
such that $\partial f:N_0\rightarrow N_1$ is an isometry.
Then
 \begin{equation}\label{eee1}Det(\Phi,f)\epsilon(M_0,N_0,F_0)
(g^{N_0})=\epsilon(M_1,N_1,F_1)(g^{N_1}).\end{equation}
Moreover,
\begin{equation}\label{eee2}Tr\:\epsilon(-M,-N,F)(g^N)\otimes
\epsilon(M,N,F)(g^N)=(-1)^d,\end{equation}
where $2d=dim\:ker(D_{F_N})+dim(F)dim\:ker(D_N)$.

\section{The definition of $H$ and $E$}

In the last section we introduced a difference construction
in order to make things independent of the Riemannian metric
in the interior
of $(M,N)$. We will use a categorial construction in order to define
an element
of a line associated with the boundary, which is independent
of the Riemannian metric of $N$.

Consider the category $Met(N,F)$. The objects of $Met(N,F)$ are the
Riemannian
metrics on $N$.
For each pair of metrics $g^N_0,g^N_1$ there is exactly one morphism
$g^N_0\rightarrow g^N_1$ (a formal object).
Recall, that $W$ denotes the category of finite-dimensional complex
Hilbert
spaces.
We define a functor $Z:Met(N,F)\rightarrow W$. Let
$Z(g^N):=Det(N,F)(g^N)$.
If $g^N_0\rightarrow g^N_1$ is a morphism, then we choose some path
$g_t^N$ of
metrics connecting $g_0^N$ and $g_1^N$. We require $g_t^N$ to be
constant near
its ends. We set
\begin{eqnarray*}Z(g^N_0\rightarrow g^N_1)&=&\epsilon([0,1]\times
N,\{0\}\times
-N\cup \{1\}\times N,pr_N^\ast F)(g_0^N,g_1^N)\\&\in&
Hom(Det(N,F)(g_0^N),Det(N,F)(g_1^N)).\end{eqnarray*}
This definition is independent of the choice of the path by the
result of the
preceeding section.
The only point to be verified is the composition rule.
\begin{prop}\label{proi} Let $g_i^N$, $i=0,1,2$  be Riemannian
metrics on $N$.
Then
$$Z(g^N_0\rightarrow g^N_1\rightarrow g^N_2)=Z(g^N_1\rightarrow
g^N_2)\circ
Z(g^N_0\rightarrow g^N_1).$$
\end{prop}
{\it Proof:}
We employ the glueing formula for the $\eta$-invariant proved in
\cite{bunke936}.
Let \linebreak[4]$(M_0,N_0,F_0)$ and $(M_1,N_1,F_1)$ be oriented
Riemannian manifolds with boundary with $N_0=-N_1$ (isometrically)
and
$\Phi:F_{0|N_0}\rightarrow F_{1|N_1}$ be a given identification.
We indicated only the boundary components, where the glueing takes
place. There
may be more boundary components, which are not affected by the
glueing.
We can glue at $N$ obtaining a new manifold $M=M_0\sharp_N M_1$,
which may have
boundaries together with a new flat bundle $F\rightarrow M$.
Choose Lagrangian subspaces for every boundary component, in
particular
$L_0$ for $N_0$ and $L_1$ for $N_1$, in order to define the boundary
conditions.
Then we have proved in \cite{bunke936}
\begin{theorem}[The glueing formula for the $\eta$-invariant]
\begin{equation}\label{e1}
 \bar{\eta}(M,F)=\bar{\eta}(M_0,N_0,F_0,L_0)+\bar{\eta}
(M_1,N_1,F_1,L_1)+m(L_0,L_1)+dim(L_0\cap L_1) (mod\:\:2\Z).
\end{equation}
\end{theorem}
We apply this theorem in order to show
\begin{equation}\label{e2}Tr\: e(M_0,N_0,F_0)\otimes
e(M_1,N_1,F_1)=e(M,F)\end{equation}
(again we omit the other boundary components in the notation).
We use the same Lagrangian $L_0, L_1:=\Phi(L_0)$ in order to define
$$e(M,N_0,F_0)=e^{\pi\imath
\bar{\eta}(M_0,N_0,F_0,L_0)}det(-\imath\sigma_{L_0}^+(N_0))$$
$$e(M,N_1,F_1)=e^{\pi\imath
\bar{\eta}(M_1,N_1,F_1,L_1)}det(-\imath\sigma_{L_1}^+(N_1)).$$
Under the identification $h(\Phi):h(N_0,F_0)\rightarrow h(N_1,F_1)^v$
we have
$$Tr\: -\imath\sigma_{L_0}^+(N_0)\otimes
-\imath\sigma_{L_1}^+(N_1)=(-1)^{dim(V)/2}.$$
This sign cancels with the $e^{\pi\imath dim(L_0\cap
\Phi^{-1}(L_1))}$ of
(\ref{e1}) and $m(L_0,\Phi^{-1}(L_1))=0$.
Thus, we obtain (\ref{e2}).
Applying the same argument to the trivial bundle we obtain
\begin{kor}\label{k1}
$$Tr\:\epsilon(M_0,N_0,F_0)(g^N)\otimes
\epsilon(M_1,N_1,F_1)(g^N)=\epsilon(M,F).$$
\end{kor}
Proposition \ref{proi} immediately follows from Corollary \ref{k1}.
$\Box$\newline

The category $Met(N,F)$ is connected and the functor $Z$ has trivial
holonomy.
Consider the cylinder
$$Cyl:=([0,1]\times N,\{0\}\times -N\cup \{1\}\times N,pr_N^\ast
F,(dr)^2\oplus
g^N),$$
where we use the constant path of metrics given by $g^N$.
We have to show $$\epsilon(Cyl)(g^N,g^N)=id\in
Hom(Det(N,F),Det(N,F)).$$
This can be seen by glueing the boundary components using \ref{k1}
and
$$\epsilon(S^1\times N,pr_N^\ast F)=1.$$

Now we define the functor $H:\partial D\rightarrow W$ by
$H(N,F):=\Gamma(Z)$.
Consider a morphism $(\Phi,f):(N_0,F_0)\rightarrow (N_1,F_1)$
in the category $\partial D$. We have to define
$H(\Phi,f):H(N_0,F_0)\rightarrow H(N_1,F_1)$.
Choose a Riemannian metric $g^{N_1}$ on $N_1$ and take
$g^{N_0}:=f^\ast
g^{N_1}$.
Then $f$ becomes an isometry. The choice of the Riemannian metrics
fixes isomorphisms $i_k:H(N_k,F_k)\rightarrow Det(N_k,F_k)$, $k=0,1$.
Moreover there is a natural isomorphism
$Det(\Phi,f):Det(N_0,F_0)\rightarrow
Det(N_1,F_1)$.
Letting $H(\Phi,f):=i_1^{-1}\circ Det(\Phi,f)\circ i_0$ it is easy to
see that
this definition does not depend on the choice of the Riemannian
metric
$g^{N_1}$.

For any object $(M,N,F)$ of $D$ we will now define the vector
$E(M,N,F)\in
H(N,F)$.
Here we use the following fact.
\begin{prop}
Let $(M,N,F)\in D$.
The correspondence $$Met(N,F)\ni g^N\mapsto \epsilon(M,N,F)(g^N)\in
Z(g^N)$$
defines a section $E(M,N,F)\in H(N,F)$.
\end{prop}
{\it Proof:}
Let $g^N_0\rightarrow g_1^N$ be a morphism of $Met(N,F)$.
Let $g_t^N$ be a path between $g_0$ and $g_1$ being constant near its
ends.
Then we must verify that
$$Tr\:\epsilon(M,N,F)(g_0^N)\otimes \epsilon([0,1]\times
N,\{0\}\times -N\cup
\{1\}\times N,pr_N^\ast F)(g_0^N,g_1^N)= \epsilon(M,N,F)(g_1^N).$$
But this follows from Corollary \ref{k1}.
$\Box$\newline

It is easy to check that $(M,N,F)\mapsto E(M,N,F)$ is a section of
the functor
$H\circ\partial:D\rightarrow W$.
\begin{theorem}
The functor $H:\partial D\rightarrow W$ and the section
$E\in\Gamma(H\circ
\partial)$ define
a local Lagrangian of a TQFT on the category $D$.
\end{theorem}
{\it Proof:}
The orientation axiom (up to a sign) follows
from (\ref{eee2}). The additivity axiom is obvious and locality
follows
once more from Corollary \ref{k1}. $\Box$\newline

What we have constructed, is a classical TQFT. The next step would
be to "take sums" over the set of flat bundles. This is the so-called
second
quantization (see Freed/Quinn \cite{freedquinn93}).
This step is much more complicated.
Our theory is very similar to the TQFT's, which have already been
constructed
using the
Chern-Simons action.
Both theories have similar infinitesimal variation formulas.
While for the Chern-Simons gauge theory the values on the
closed manifolds are the values of the Chern-Simons functional of the
flat
connections,
in our case the values are the
$\xi$-invariants of Atiyah-Patodi-Singer
\cite{atiyahpatodisinger752}, which
are, in fact, also
a sort of Chern-Simons invariants.
The difference of the theory defined with the $\eta$-invariant
and the Chern-Simons theory is potentially located in its contents
for
manifolds with non-empty boundaries.
For the Chern-Simons action the second quantization has been carried
out by
Freed/Quinn
\cite{freedquinn93} for a finite gauge group.


\bibliographystyle{plain}



\end{document}